\RequirePackage{ifpdf}
\ifpdf 
\documentclass[pdftex]{sigma}
\else
\documentclass{sigma}
\fi

\begin{document}

\renewcommand{\PaperNumber}{075}

\FirstPageHeading

\ShortArticleName{The Veldkamp Space of Two-Qubits}

\ArticleName{The Veldkamp Space of Two-Qubits}

\Author{Metod SANIGA~$^\dag$, Michel PLANAT~$^\ddag$, Petr PRACNA~$^\S$ and Hans HAVLICEK~$^\P$}

\AuthorNameForHeading{M. Saniga, M. Planat, P. Pracna and H. Havlicek}

\Address{$^\dag$~Astronomical Institute, Slovak Academy of Sciences,\\
$\phantom{^\dag}$~SK-05960 Tatransk\' a Lomnica, Slovak Republic}
\EmailD{\href{mailto:msaniga@astro.sk}{msaniga@astro.sk}}
\URLaddressD{\url{http://www.ta3.sk/~msaniga/}}

\Address{$^\ddag$~Institut FEMTO-ST, CNRS, D\' epartement LPMO,
32 Avenue de
l'Observatoire,\\ 
$\phantom{^\dag}$~F-25044 Besan\c con Cedex, France}
\EmailD{\href{mailto:michel.planat@femto-st.fr}{michel.planat@femto-st.fr}}

\Address{$^\S$~J. Heyrovsk\' y Institute of Physical
Chemistry, Academy of Sciences of the Czech Republic,\\ 
$\phantom{^\S}$~Dolej\v skova 3, CZ-182 23 Prague 8, Czech
Republic}
\EmailD{\href{mailto:pracna@jh-inst.cas.cz}{pracna@jh-inst.cas.cz}}

\Address{$^\P$~Institut f\" ur Diskrete Mathematik und Geometrie,
Technische Universit\" at Wien,\\
$\phantom{^\P}$~Wiedner Hauptstra\ss e 8--10, A-1040 Vienna, Austria}
\EmailD{\href{mailto:havlicek@geometrie.tuwien.ac.at}{havlicek@geometrie.tuwien.ac.at}}

\ArticleDates{Received April 13, 2007, in f\/inal form June 18, 2007; Published online June 29, 2007}

\Abstract{Given a remarkable representation of the generalized
Pauli operators of two-qubits in terms of the points of the
generalized quadrangle of order two, $W(2)$, it is shown that
specif\/ic subsets of these operators can also be associated with
the points and lines of the four-dimensional projective space over
the Galois f\/ield with two elements -- the so-called Veldkamp
space of $W(2)$. An intriguing novelty is the recognition of (uni- and tri-centric) triads
and specif\/ic pentads of the Pauli operators in addition to the ``classical" subsets
answering to geometric hyperplanes of $W(2)$.}

\Keywords{generalized quadrangles; Veldkamp spaces; Pauli operators of two-qubits}

\Classification{51Exx; 81R99}

\vspace{-4mm}
\section{Introduction}
A deeper understanding of the structure of Hilbert spaces of
f\/inite dimensions is of utmost importance for quantum information
theory. Recently, we made an important step in this respect by
demonstrating that the commutation algebra of the generalized Pauli operators
on the $2^N$-dimensional Hilbert spaces is embodied in the
geometry of the symplectic polar space of rank $N$ and order two
\cite{saplpr,sapl,plsa}. The case of two-qubit operator space,
$N=2$, was scrutinized in very detail \cite{saplpr,plsa} by explicitly demonstrating,
in dif\/ferent ways, the correspondence between various subsets of
the generalized Pauli operators/matrices and the fundamental
subgeometries of the associated rank-two polar space -- the
(unique) generalized quadrangle of order two. In
this paper we will reveal another interesting geometry hidden
behind the Pauli operators of two-qubits, namely that of the
Veldkamp space def\/ined on this generalized quadrangle.

\section{Finite generalized quadrangles and Veldkamp spaces}

{\samepage In this section we will brief\/ly highlight the basics of the theory of f\/inite generalized
quadrangles~\cite{paythas} and introduce the concept of the Veldkamp space of a point-line incidence geometry~\cite{buek}
to be employed in what follows.}

A {\it finite generalized quadrangle} of order $(s, t)$, usually
denoted GQ($s, t$), is an incidence structure $S = (P, B, {\rm
I})$, where $P$ and $B$ are disjoint (non-empty) sets of objects,
called respectively points and lines, and where I is a symmetric
point-line incidence relation satisfying the following axioms
\cite{paythas}: (i) each point is incident with $1 + t$ lines ($t
\geq 1$) and two distinct points are incident with at most one
line; (ii) each line is incident with $1 + s$ points ($s \geq 1$)
and two distinct lines are incident with at most one point;  and
(iii) if $x$ is a point and $L$ is a line not incident with $x$,
then there exists a unique pair $(y, M) \in  P \times B$ for which
$x {\rm I} M {\rm I} y {\rm I} L$; from these axioms it readily
follows that $|P| = (s+1)(st+1)$ and $|B| = (t+1)(st+1)$. It is obvious that there exists
a point-line duality with respect to which each of the axioms is self-dual.
Interchanging points and lines in $S$ thus yields a generalized
quadrangle $S^{D}$ of order $(t, s)$, called the dual of $S$. If
$s = t$, $S$ is said to have order $s$. The generalized quadrangle
of order $(s, 1)$ is called a grid and that of order $(1, t)$ a
dual grid. A generalized quadrangle with both $s > 1$ and $t > 1$
is called thick.

     Given two points $x$ and $y$ of $S$ one writes $x \sim y$ and says that $x$ and $y$ are collinear if
     there exists a line $L$ of $S$ incident with both. For any $x \in P$ denote $x^{\perp} = \{y \in P | y \sim x \}$
     and note that $x \in x^{\perp}$;  obviously, $x^{\perp} = 1+s+st$. 
     Given an arbitrary subset $A$ of $P$, the {\it perp}(-set) of~$A$,~$A^{\perp}$, 
     is def\/ined as $A^{\perp} =  \bigcap \{x^{\perp} | x \in A\}$ and
     $A^{\perp \perp} := (A^{\perp})^{\perp}$. A triple of pairwise non-collinear points of $S$ is called a {\it triad}; given any triad $T$, a point of $T^{\perp}$ is called its center and we say that $T$ is acentric, centric or unicentric according as $|T^{\perp}|$ is, respectively, zero, non-zero or one. An ovoid of a generalized quadrangle $S$ is a set of points of $S$ such that each line of $S$ is incident with exactly one point of the set;  hence, each ovoid contains $st + 1$ points.

The concept of crucial importance is a {\it geometric hyperplane} $H$ of a point-line geometry $\Gamma (P,B)$, which is a proper subset of $P$ such that each line of $\Gamma$ meets $H$ in one or all points \cite{ron}. For $\Gamma =$ GQ($s, t$), it is well known that $H$ is one of the following three kinds: (i) the perp-set of a point $x$,  $x^{\perp}$; (ii) a (full) subquadrangle of order ($s,t'$), $t' < t$; and (iii) an ovoid.

     Finally, we need to introduce the notion of the {\it Veldkamp space} 
     of a point-line incidence geometry $\Gamma(P,B)$, $\mathcal{V}(\Gamma)$ \cite{buek}. 
     $\mathcal{V}(\Gamma)$  is the space in  which (i) a point is a geometric 
     hyperplane of~$\Gamma$ and (ii)  a line is the collection $H_{1}H_{2}$ of all geometric hyperplanes $H$ of $\Gamma$  such that $H_{1} \bigcap H_{2} = H_{1} \bigcap H = H_{2} \bigcap H$ or $H = H_{i}$ ($i = 1, 2$), where $H_{1}$ and  $H_{ 2}$ are  distinct points of $\mathcal{V}(\Gamma)$.\footnote{It is important to mention here that
     the def\/inition of Veldkamp space given by Shult in \cite{shult} is more restrictive than that of Buekenhout and Cohen \cite{buek} adopted in this paper.}  If  $\Gamma = S$, from the preceding paragraph we learn that the points of  $\mathcal{V}(S)$ are, in general, of three dif\/ferent types.

\begin{figure}[t]
\centerline{\includegraphics{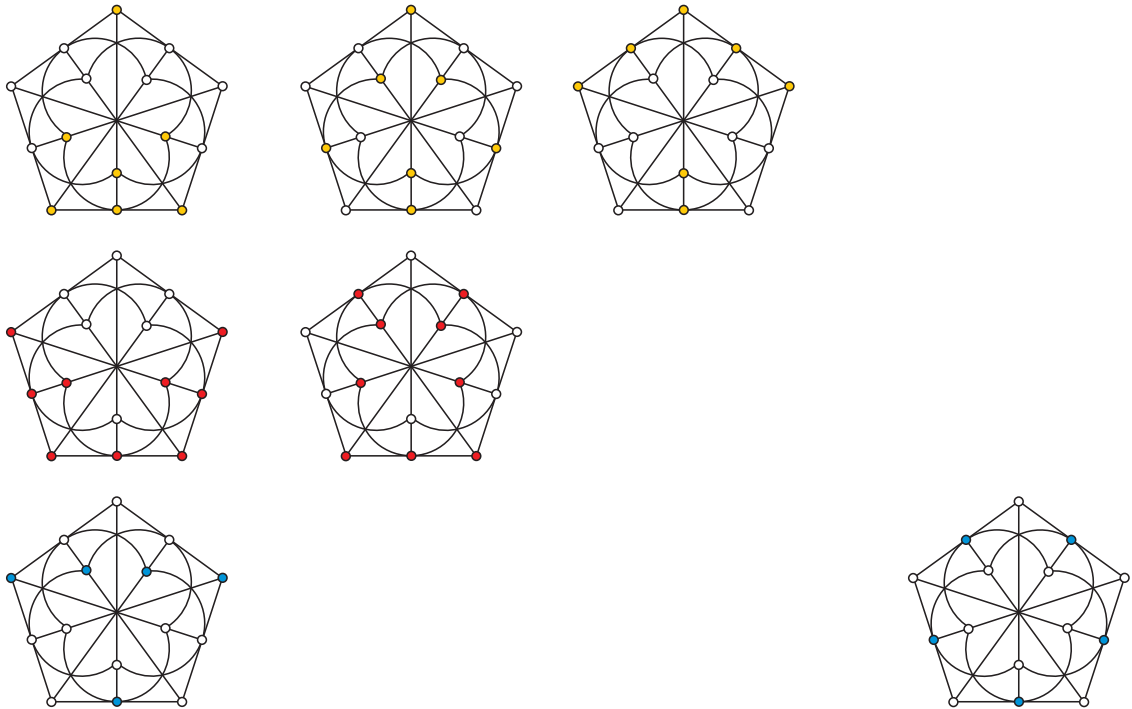}}
\caption{The three kinds of geometric hyperplanes of $W(2)$. 
The points of the quadrangle are represented by small 
circles and its lines are illustrated by the straight segments as
well as by the segments of circles; note that not 
every intersection of two segments counts for a point of
the quadrangle. The upper panel shows the points' perp-sets (yellow bullets), 
the middle panel grids (red bullets) and the bottom panel ovoids (blue bullets); the use
of dif\/ferent colouring will become clear later. 
Each picture~-- except that in the bottom right-hand corner~--
stands for f\/ive dif\/ferent hyperplanes, the four other being obtained 
from it by its successive rotations through 72 degrees around the center of the pentagon.}
\vspace{-1mm}
\end{figure}

\begin{figure}[t]
\centerline{\includegraphics[width=7.cm,height=7.cm,clip=]{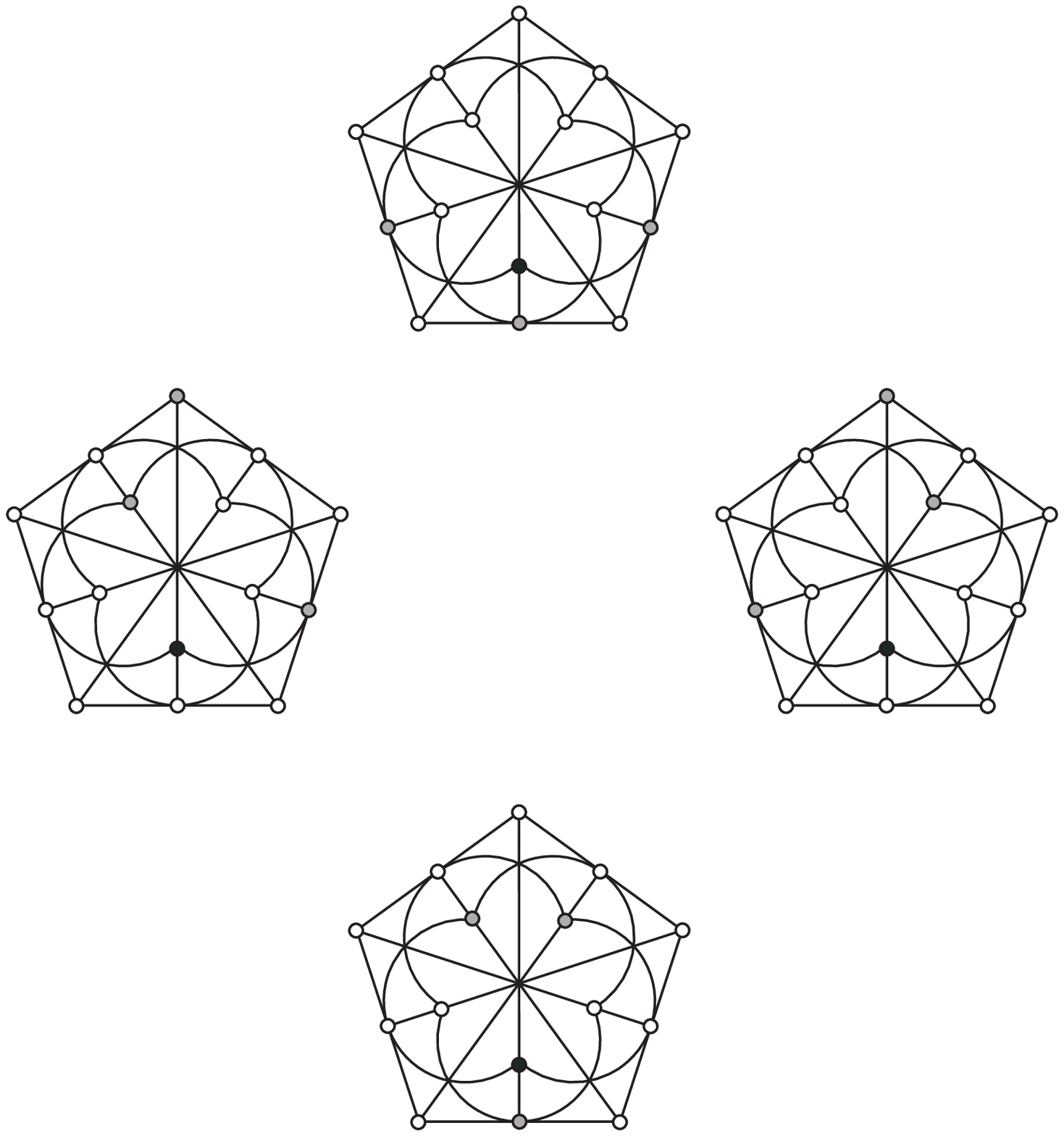}
\includegraphics[width=7.cm,height=7.cm,clip=]{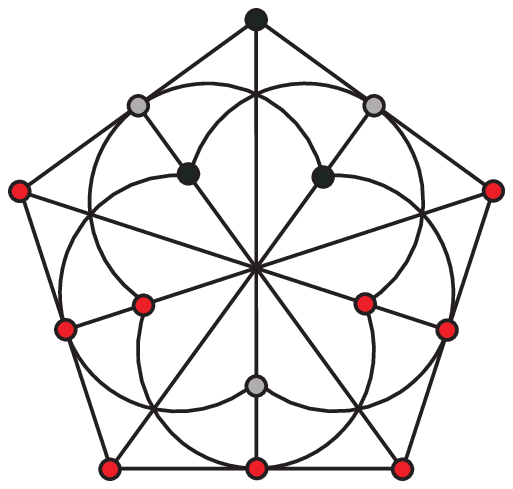}}
 \caption{{\it Left}: -- The four distinct
unicentric triads (grey bullets) and their common center (black
bullet); note that the triads intersect pairwise in a single point
and their union covers fully the center's perp-set. 
{\it Right}:~-- A grid (red bullets) and its complement as a disjoint union of
two complementary tricentric triads (black and grey bullets); the two triads are also seen
to comprise a dual grid (of order ($1,2$)).}\vspace{-3mm}
\end{figure}

\begin{figure}[t]
\centerline{\includegraphics[width=13.2cm,clip=]{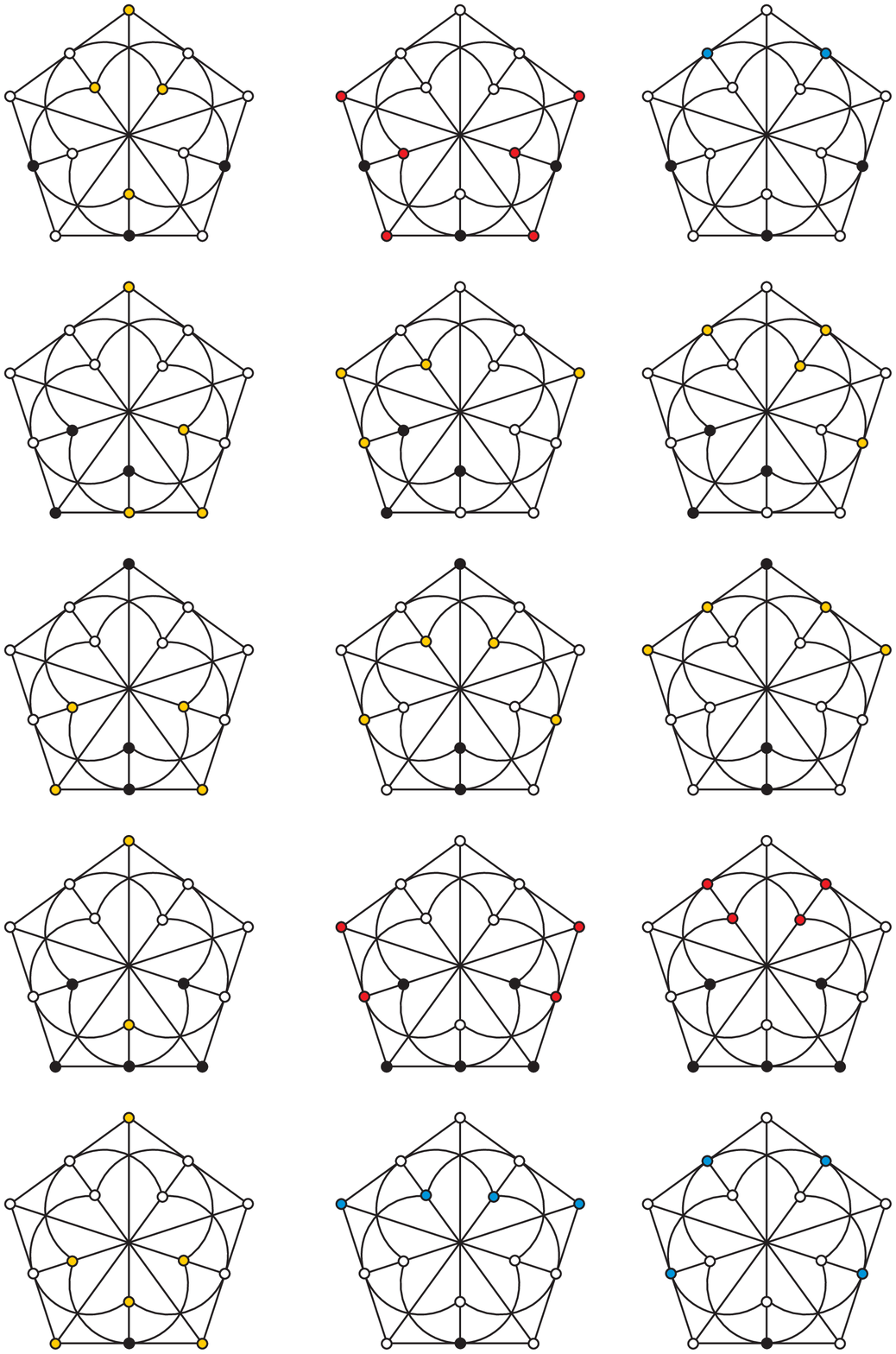}}
 \caption{The f\/ive dif\/ferent kinds of the lines of
$\mathcal{V}(W(2))$, each being uniquely determined by the
properties of its core-set (black bullets). Note that the
``yellow'' hyperplanes (i.e., perp-sets) occur in each type, and yellow
is also the colour of two homogeneous (i.e., endowed with only one
kind of a~hyperplane) types (2nd and 3rd row).  It is also worth
mentioning that the cardinality of core-sets is an odd number not
exceeding f\/ive. The three hyperplanes of any line are always in
such relation to each other that their union comprises all the
points of $W(2)$.}
\end{figure}

\section{The smallest thick GQ and its Veldkamp space}

The smallest thick GQ is obviously the one with $s=t=2$, dubbed
the ``doily." This quadrangle has a number of interesting
representations of which we mention the most important two
\cite{paythas}. One, frequently denoted as $W_{3}(2)$ or simply
$W(2)$, is in terms of the points of $PG(3,2)$ (i.e., the
three-dimensional projective space over the Galois f\/ield with two
elements) together with the totally isotropic lines with respect
to a symplectic polarity. The other, usually denoted as $Q(4,2)$,
is in terms of points and lines of a parabolic quadric in
$PG(4,2)$. By abuse of notation, any GQ isomorphic to $W(2)$ will
also be denoted by this symbol. From the preceding section we
readily get that $W(2)$ is endowed with 15~points/lines, each line
contains three points and, dually, each point is on three lines;
moreover, it is a self-dual object, i.e., isomorphic to its
dual. $W(2)$ features all the three kinds of hyperplanes, of the
following cardinalities~\cite{buek}: 15~perp-sets, $x^{\perp}$,
seven points each; 10 grids (of order $(2,1)$), nine points each;
and six ovoids, f\/ive points each~-- as depicted in Fig.~1. The
quadrangle exhibits two distinct kinds of triads, viz. unicentric
and tricentric. A point of $W(2)$ is the center of four distinct
unicentric triads (Fig.~2, left); hence, the number of such
triads is $4 \times 15 = 60$. Tricentric triads always come in
``complementary'' pairs, one representing the centers of the other,
and each such pair is the complement of a~grid of $W(2)$ (Fig.~2, 
right); hence, the number of such triads is $2 \times 10 = 20$.
A unicentric triad is always a~subset of an ovoid, which is never
the case for a tricentric triad; the latter, in
graph-combinatorial terms, representing a complete bipartite graph
on six vertices. Now, we have enough background information at
hand to reveal the structure of the Veldkamp space of our
``doily'', $\mathcal{V}(W(2))$.\footnote{As this paper is primarily
aimed at physicists rather than mathematicians, in what follows we
opt for an elementary and self-contained exposition of the
Veldkamp space of $W(2)$; this explanation is based only upon some
very simple properties of $W(2)$ readily to be grasped from its
depiction as ``the doily'', and does not presuppose/require any
further background from the reader.}

From the def\/inition given in Section~2, we easily see that  $\mathcal{V}(W(2))$ consists of 31 points of which
15 are represented/generated by single-point perp-sets, 10 by grids and six by ovoids. The lines of $\mathcal{V}(W(2))$
feature three points each and
are of f\/ive distinct types, as illustrated in Fig.~3.  These types dif\/fer from each other in the cardinality and
structure of ``core-sets'', i.e., the sets of points of $W(2)$ shared by all the three hyperplanes forming
a given line. As it is obvious from Fig.~3, the lines of the f\/irst 
three types (the f\/irst three rows of the f\/igure) 
have the core-sets of the same cardinality, three, 
dif\/fering from each other only in the structure of 
these sets as being unicentric triads, tricentric 
triads and triples of collinear points, respectively. 
The lines of the fourth type have as core-sets pentads 
of points, each being a quadruple of points collinear with a given point of $W(2)$,
whereas core-sets of the last type's lines 
feature just a~single point. A much more interesting issue is 
the composition of the lines. Just a~brief look at Fig.~3 reveals that 
geometric hyperplanes of only one kind, namely perp-sets, are present on 
each line of $\mathcal{V}(W(2))$;  grids and ovoids occur only on two kinds of the lines.
We also see that the purely homogeneous types are those whose core-sets feature collinear triples and tricentric
triads, the most heterogeneous type~-- the one exhibiting all the three kinds of hyperplanes~-- 
being that characterized by unicentric triads. We also notice that there are no 
lines comprising solely grids and/or solely ovoids, nor the lines featuring 
only grids and ovoids, which seems to be connected with the fact 
that the cardinality of a core-set is an odd number. 
From the properties
of $W(2)$ and its triads as discussed above it readily 
follows that the number of the lines of type one to 
f\/ive is 60, 20, 15, 45 and 15, respectively, totalling 155. 
All these observations and facts are gathered in Table~1.
We conclude this section with the observation that
$\mathcal{V}(W(2))$ has the same number of points (31) 
and lines (155) as $PG(4,2)$, the four-dimensional 
projective space over the Galois f\/ield of two elements~\cite{hirthas}; 
this is not a coincidence, as the two spaces are, in fact, isomorphic to each other~\cite{buek}.

\section[Pauli operators of two-qubits in light of $\mathcal{V}(W(2))$]{Pauli 
operators of two-qubits in light of $\boldsymbol{\mathcal{V}(W(2))}$}

As discovered in \cite{saplpr} (see also \cite{plsa}), the f\/ifteen generalized Pauli
operators/matrices associated with the Hilbert space of two-qubits
(see, e.g., \cite{lbz}) can be put into a one-to-one
correspondence with the f\/ifteen points of the generalized quadrangle
$W(2)$ in such a way that their commutation algebra is completely
and uniquely reproduced by the geometry of $W(2)$ in which the
concept commuting/non-commuting translates into that of
collinear/non-collinear. Given this mapping, it was possible to
ascribe a def\/initive geometrical meaning to sets of three pairwise
commuting generalized Pauli operators in terms of lines of $W(2)$
and to other three kinds of distinguished subsets of the operators
having their counterparts  in geometric hyperplanes of $W(2)$ as
shown in Table~2 (see \cite{saplpr,plsa} for more details). Yet,
$\mathcal{V}(W(2))$ puts this bijection in a dif\/ferent light, in
which other three subsets of the Pauli operators come into play,
namely those represented by the two types of a triad and by the
specif\/ic pentads occurring as the core-sets of the lines of
$\mathcal{V}(W(2))$ (Table~1). As already mentioned, the role of
tricentric triads of the operators has been recognized in disguise
of complete bipartite graphs on six vertices \cite{plsa}. A true
novelty here is obviously {\it uni}centric triads and {\it
pentads} of the generalized Pauli operators as these are all
intimately connected with single-point perp-sets; given a point of
$W(2)$ (i.e., a generalized Pauli operator of two-qubits), its
perp-set fully encompasses {\it four} unicentric triads (Fig.~2,
left) and {\it three} pentads (Fig.~3, 4th row) of
points/operators. This feature has also a very interesting aspect
in connection with the conjecture relating the existence of
mutually unbiased bases and f\/inite projective planes raised in
\cite{spr}, because with each point $x$ of $W(2)$ there is
associated a~projective plane of order two (the Fano plane) whose
points are the elements of $x^{\perp}$ and whose lines are the
spans $\{u,v\}^{\perp \perp}$, where $u,v \in x^{\perp}$ with $u
\neq v$ \cite{paythas}.

\begin{table}[t]
\begin{center}\small
\caption{A succinct summary of the properties of the f\/ive dif\/ferent types of the lines of $\mathcal{V}(W(2))$
in terms of the core-sets and the types of geometric hyperplanes featured by a generic line of a given type.
The last column gives the total number of lines per the corresponding type.} \vspace*{0.2cm}
 \begin{tabular}{|l|ccc|c|}
\hline \hline
Type of Core-Set & Perp-Sets & Grids & Ovoids & $\#$ \\
\hline
Single Point & 1 &  0 & 2 & 15 \\
Collinear Triple & 3 & 0 & 0 & 15 \\
Unicentric Triad & 1 & 1 &  1 & 60 \\
Tricentric Triad & 3 & 0 &  0 & 20 \\
Pentad & 1 & 2 & 0 & 45\\
\hline \hline
\end{tabular}
\end{center}
\vspace{-5mm}
\end{table}

\begin{table}[t]\small
\begin{center}
\caption{Three kinds of the distinguished subsets of the
generalized Pauli operators of two-qubits (PO) viewed as the geometric
hyperplanes in the generalized quadrangle of order two (GQ) \cite{saplpr,plsa}.}
\vspace*{0.2cm}
\begin{tabular}{llll}
\hline \hline
\vspace*{-.3cm} \\
PO  & set of f\/ive mutually    & set of six operators  & nine operators of \\
   & non-commuting operators  & commuting with a given one &  a Mermin's square\\
GQ & ovoid  &   perp-set$\setminus$\{reference point\} &   grid \\
\vspace*{-.3cm} \\
\hline \hline
\end{tabular}
\vspace{-5mm}
\end{center}
\end{table}

Identifying the Pauli operators of a two-qubit system with the 
points of the generalized quadrangle of order two led to the 
discovery of three distinguished subsets of the operators in terms 
of geometric hyperplanes of the quadrangle. Here we go one level 
higher, and identifying these subsets with the points of the associated 
Veldkamp space leads to recognition of nother remarkable subsets of 
the Pauli operators, viz. unicentric triads and pentads. It is really 
intriguing to see that these are the core-sets of the two kinds of lines 
that both feature {\it grids} alias {\it Mermin squares}. As it is well 
known, Mermin squares, which reveal certain important aspects of the 
entanglement of the system, play a crucial role in the proof of the 
Kochen--Specker theorem in dimension four and 
our approach gives a novel geometrical meaning to this \cite{plsa,thmp}. 
At the Veldkamp space level it turns out of particular importance to study 
relations between eigenvectors of the above-mentioned unicentric triads and 
pentad of operators in order to reveal f\/iner, hitherto unnoticed traits of the 
structure of Mermin squares. These seem to be intimately connected with the existence 
of {\it outer} automorphisms of the symmetric group on six letters, which is the 
full group of automorphisms of our quadrangle; as this 
group is the only symmetric group possessing (non-trivial) 
outer automorphisms, this implies that two-qubits have 
a rather special footing among multiple qubit systems. 
All these aspects deserve special attention and will therefore be dealt with in a separate paper.

Concerning three-qubits, our preliminary study indicates 
that the corresponding f\/inite geo\-metry dif\/fers fundamentally 
from that of $W(2)$ in the sense that it contains multi-lines, i.e., 
two or more lines passing through two distinct points~\cite{spie}. 
As we do not have a full picture at hand yet, we cannot see if 
it admits hyperplanes and so lends itself to constructing 
the corresponding Veldkamp space. If the latter does exist, 
it is likely to dif\/fer substantially from that of two-qubits, 
which would imply the expected dif\/ference between entanglement 
properties of the two kinds of systems; if not, this will only 
further strengthen the above-mentioned uniqueness of two-qubits.

\section{Conclusion}
By employing the concept of the Veldkamp space of the generalized quadrangle of order two, we were able
to recognize other, on top of those examined in \cite{saplpr,sapl,plsa}, 
distinguished subsets of generalized Pauli operators of two-level quantum systems, 
namely unicentric triads and pentads of them. It may well be that these 
two kinds of subsets of the two-qubit Pauli operators hold an important key for 
getting deeper insights into the nature of  f\/inite geometries underlying multiple 
higher-level quantum systems \cite{spie,koen}, in particular
when the dimension of Hilbert space is not a power of a prime \cite{koen2}.

\subsection*{Acknowledgements}

This work was partially supported by the Science and
Technology Assistance Agency under the contract $\#$
APVT--51--012704, the VEGA grant agency projects $\#$ 2/6070/26
and $\#$ 7012 (all from Slovak Republic), the trans-national
ECO-NET project $\#$ 12651NJ ``Geometries Over Finite Rings and
the Properties of Mutually Unbiased Bases" (France), the
CNRS--SAV Project $\#$ 20246 ``Projective and Related Geometries
for Quantum Information" (France/Slovakia) and by the $\langle$Action Austria--Slovakia$\rangle$
project $\#$ 58s2 ``Finite Geometries Behind Hilbert Spaces".

\pdfbookmark[1]{References}{ref}

\end{document}